\documentclass[pdflatex,sn-mathphys]{sn-jnl}

\jyear{2022}%

\theoremstyle{thmstyleone}%

\theoremstyle{thmstyletwo}%

\theoremstyle{thmstylethree}%

\begin{document}

\title[The Point of Primitive Ontology]{The Point of Primitive Ontology}

\author*[1]{\fnm{Dustin} \sur{Lazarovici}}\email{dustin@technion.ac.il}

\author*[1,2]{\fnm{Paula} \sur{Reichert}}\email{reichert@math.lmu.de}

\affil[1]{\orgdiv{Department of Humanities and Arts}, \orgname{Technion -- Israel Institute of Technology}, \orgaddress{\city{Haifa}, \country{Israel}}}

\affil[2]{\orgdiv{Mathematisches Institut}, \orgname{Ludwig-Maximilians-Universität München}, \country{Germany}}



\abstract{Bohmian mechanics grounds the predictions of quantum mechanics in precise dynamical laws for a primitive ontology of point particles. In an appraisal of the de-Broglie-Bohm theory, the paper discusses the crucial epistemological and conceptual role that a primitive ontology plays within a physical theory. It argues that quantum theories without primitive ontology fail to make contact with observable reality in a clear and consistent manner. Finally, it discusses Einstein's epistemological model and why it supports the primitive ontology approach.}


\maketitle
\section{Introduction}
 One of the main virtues of the physical theory that we now call Bohmian mechanics\footnote{For modern presentations of the theory, see \cite{durr.teufel2009, duerr.etal2013, bricmont2016, durr.lazarovici2020}.} and that goes back to the 1952 work of Bohm \cite{bohm1952a} and 1927 ideas of de Broglie  \cite{debroglie1927a,deBroglie1927b} is that it could be true. By this, we are not referring to the fact that the theory accounts for all known phenomena of non-relativistic quantum physics (although it does). Nor are we suggesting that Bohmian mechanics might be the fundamental theory of our universe (it is not, at least in its current form, but see \cite{durr.etal2013, durr.etal2013a} for possible generalizations to relativistic quantum mechanics and QFT). We mean that Bohmian mechanics makes precise, objective, and unambiguous statements about the physical world, statements that are semantically capable of being true. For this reason, Bohmian mechanics is sometimes referred to as a ``realistic'' quantum theory. This is, however, a misnomer. Scientific realism (which is apparently at stake here) is not a feature that a physical theory may or may not possess but a philosophical attitude that one may or may not take towards a theory. That one \emph{can} be a realist about Bohmian mechanics means precisely that the theory is of a form and shape that allows it to be true (or approximately so). And this distinguishes the Bohmian theory from most so-called ``interpretations'' of quantum mechanics and, above all, from the bare formalism of textbook quantum mechanics, the loose set of rules and computational techniques that Maudlin \cite{maudlin2019} describes as the ``quantum recipe.''

Bohmian mechanics is a theory about the motion of point particles postulated as the basic constituents of matter in space and time. As such, the theory is (approximately) true if the particles succeed in referring to something in the world that behaves (at least approximately and in appropriate regimes) according to the Bohmian laws. We can empirically test the theory by comparing its predictions about the spatio-temporal distribution of matter with observations, including the statistical distribution of pointer positions, display readings, detector clicks\footnote{Even the sound of a clicking detector corresponds to matter in motion (sound waves in air), and something in the configuration of the detector has to change to produce them.}, or whatever else records the outcomes of measurement experiments. An anti-realist may be skeptical, agnostic, or indifferent about the truth of the theory but still appreciate that it provides a clear, consistent, and successful systematization of the phenomena, as well as a coherent ``image'' of the world that goes along with it. 

\section{Ontology and Epistemology}

Bohmian mechanics, in other words, is a theory about what there is and how it behaves. The what-there-is part is the \emph{ontology} of the theory. The how-it-behaves part are the laws of motion, given by two precise mathematical equations. More specifically, the Bohmian theory describes (3+1)-dimensional worlds populated by matter, including the kind of macro-objects that we perceive around us. We can analyze the theory and find that typical solutions look like the world we experience, manifesting, in particular, the statistical regularities predicted by standard quantum mechanics and confirmed by experiments. This, in a nutshell, is the \emph{epistemology} of the theory, the way it connects to and is tested against observable facts. 

This general structure is pretty much analogous for Newton's theory and Maxwell's theory and Einstein's theory, as Bohm and Hiley already stressed.

\begin{quote}
  In classical physics there was never a serious problem either about the ontology, or about the epistemology. With regard to the ontology, one assumed the existence of particles and fields which were taken to be essentially independent of the human observer. The epistemology was then almost self-evident because the observing apparatus was supposed to obey the same objective laws as the observed system, so that the measurement process could be understood as a special case of the general laws applying to the entire universe. \citep[p. 13]{bohm1993} 
\end{quote}

That the same is true in Bohm's theory is only remarkable because the orthodoxy had insisted (and often keeps insisting) that quantum physics cannot be like this. It is only offensive because this insistence has licensed the wildest claims about the bizarreness or fuzziness or subjectivity or incomprehensibility of the quantum realm that the mere existence of Bohmian mechanics reveals as not necessarily wrong but gratuitous.

Standard quantum mechanics has no clear ontology but wants to speak directly (and exclusively) about ``observable values'' or ``measurement outcomes''. Consequently, it invokes instead, or in addition to, precise dynamical equations (viz. the Schrödinger equation) various rules and postulates about ``observables'' and their ``measurements.''  The problem with this approach is not that it leaves some realist itch unscratched but that it is a bad way of doing physics. It is not surprising that such rules can be ambiguous and even inconsistent when applied outside the context of more or less controlled and familiar experiments. The concept of measurements is vague, to begin with, and often refers to complex processes that must be analyzable in physical terms and cannot be treated as primitive in the formulation of a fundamental theory. More basically, one may wonder if there is even reason to expect logical consistency from a formalism that does not at least purport to provide models of objective reality. (Bohr's principle of complementarity can be understood as an admission that there is not.)

Some quantum theories make an ontological commitment to the wave function (or whatever the wave function represents), which is, after all, the only object for which standard quantum mechanics provides a precise dynamical equation. They are thus theories \emph{about} the wave function (or some more abstract quantum state) but face a problem on the side of the epistemology since it is not at all obvious what a complex-valued field on a roughly $10^{80}$-dimensional space -- or some vector in an abstract Hilbert space -- has to do with the physical world that we experience. We will discuss the strategies employed to meet this challenge in Section \ref{sec:WFrealism}.

What Bohmian mechanics has that such theories lack is what John Bell described as \emph{local beables} \citep[Ch. 7]{bell2004} and what is now more commonly referred to as  a \emph{primitive ontology}.\footnote{The term \emph{primitive ontology} was first introduced by D\"urr, Goldstein, and Zangh\`i in 1992 \citep{duerr.etal1992}. For detailed discussions, see, e.g., \cite{allori.etal2008, allori2013, esfeld2020a}; cf. Maudlin's notion of \emph{primary ontology} in \cite{maudlin2013}.}  Those are localized entities in three-dimensional space or four-dimensional spacetime (or some space that coarse-grains, in an obvious sense, to the (3+1)-dimensional geometry of our experience) that the theory postulates as the microscopic constituents of matter. \emph{Stuff} like point particles, i.e., \emph{atoms} in the original sense, that could compose tables and cats and computers and measurement devices and that allows us to make empirical predictions by analyzing how this stuff moves.

The primitive ontology (PO) of a theory is conceptually and epistemically fundamental. By \emph{conceptually fundamental} we mean that the PO is not read off or interpreted into a mathematical formalism but posited as that to which the theoretical formalism ultimately refers. Other objects or structures appearing in the theory are understood through the role they play in the dynamics of the PO. This applies, in particular, to the wave function in Bohmian mechanics but also, for instance, to the gravitational potential in Newtonian theory or the electromagnetic field in classical electrodynamics (cf. \cite{lazarovici2018}). 

There are lively philosophical debates about whether the Bohmian wave function should be interpreted as a physical entity over and above the particles -- as part of the physical ontology, though not the primitive one -- or rather as part of the physical laws (see Chen  \cite{chen2019} for a good overview). But this is the kind of debate that one can leave to philosophers in good conscience; it has no more bearing on the physics than the analogous question about the Newtonian gravitational ``field''. The meaning of the wave function within Bohmian theory is clear; what this object does and how it manifests empirically, how the wave functions of subsystems are defined, and why they are an objective part of their physical state description. 

By \emph{epistemically fundamental} we mean that the PO provides the points of closest contact between theory and experience -- between the scientific and manifest images of the world as Maudlin \cite{maudlin1997} puts it, borrowing terminology from Sellars \cite{sellars1997} --- and that it is properly basic in filling this role. Nothing more can or needs to be said about what a pointer-shaped configuration of particles has to do with an actual pointer on a measurement device, at least not as a starting point for a physical analysis that looks at whether the dynamics of the particle configuration are consistent with the pointer's observed behavior. The same would be true for appropriate configurations of a field in physical space or other local beables that a theory might postulate but not, e.g., for the amplitudes and phases of a septillion-dimensional wave function or the spectrum of a self-adjoint operator on Hilbert space.

It is usually natural to also interpret the PO as \emph{metaphysically fundamental}, which means that (under a realist reading of the theory) the PO exists ``in itself,'' not grounded in or by virtue of anything else. Certainly, the existence of stones and tables and cats, etc. depends on the existence of their microscopic constituents and not the other way around. But much deeper metaphysical considerations are not essential to the concept of a primitive ontology, and it is not incoherent to think that primitive and fundamental ontology go apart. In some theories\footnote{E.g., the GRW flash theory (see below) or certain formulations of quantum field theory.} the role of the dynamical structure will include the creation of the primitive entities. The latter remain the basic building blocks of (visible) matter but their existence becomes contingent on other quantities, putting their metaphysical fundamentality into question. This illustrates that the insistence on a primitive ontology does not amount to armchair metaphysics. Physics might well discover more fundamental structures or entities, but the PO is what ultimately provides the link to observable reality. In fact, to the extent that one wants to draw a distinction between physics and metaphysics, PO is primarily a physical concept. It is an answer to two closely related questions which fall first and foremost to the physical theory: ``What is matter?'' and ``How does the theory make contact with the physical world?''.



We are nonetheless going to end the section on a philosophical note by observing that the three notions of fundamentality just discussed -- conceptual, epistemical, and ontological -- parallel those recognized in Aristotle's \emph{metaphysics} \cite[book Z1]{aristotle}. Aristotle seeks for the different ways in which substance (\textit{ousia}) is primary to all being (\textit{to on}) and finds a threefold priority: substance is primary in definition (\textit{log\^{o}i}), in recognition (\textit{gn\^{o}sei}), and in time (\textit{chron\^{o}i}). The latter,  generally understood as an ontological priority, has become the focus of the recent ``Aristotelian turn'' in  metaphysics, but it won't be the focus of this essay. To understand the point of a primitive ontology, it is more important to appreciate the central epistemic and conceptual role that it assumes within a physical theory. %


\section{The Measurement Problem}\label{sec:MP}

If one wants to say what is wrong with standard quantum mechanics, one must inevitably talk about the measurement problem. The measurement problem is essentially Schrödinger's cat paradox but its most rigorous formulation is due to Maudlin \citep{maudlin1995}. It is the logical inconsistency of the following three premises:

\begin{enumerate}
	\item The wave function $\psi$ of a system provides a complete description of its physical state, i.e., it specifies (directly or indirectly) all of the physical properties of the system.
	
	\item The wave-function always evolves in accord with
a linear dynamical equation (e.g., the Schrödinger equation).
	
	\item Measurements (usually) have definite outcomes, e.g., at the end of Schrödinger's cat experiment, the cat is either dead or alive. 
\end{enumerate}

\noindent Since the conjunction of 1--3 leads to a contradiction, any consistent formulation of quantum mechanics has to reject at least one of the premises. 

If we reject 1, i.e., that the wave function of a system provides a complete description of its physical state, we have to specify additional variables that complete the state description. The most straightforward realization of this solution is Bohmian mechanics, the ``obvious ontology evolving the obvious way'' as Sheldon Goldstein, one of its foremost advocates, puts it (private communication).

If we reject assumption 2, i.e., that the evolution of the wave function is described by a  linear (Schrödinger) equation, we have to specify what equation describes it instead. This leads to the class of \emph{objective collapse theories} whose prototype is known as GRW\index{GRW theory} (after Ghirardi, Rimini, and Weber \cite{ghirardi.etal1986}, who were the first to propose such a theory in 1986).

Rejecting assumption 3 and admitting that measurements have, in general, no definite outcome leads to \emph{Many-Worlds theories} that accept macroscopic superpositions -- which ultimately come to encompass the entire universe -- as describing the coexistence of equally ``real'' states. 

These three theories, or classes of theories, are also called \emph{quantum theories without observers} (after Bell \cite[p. 173]{bell2004}) because they provide an objective physical description in which ``the observer'' assumes no a priori distinguished role. They are not entirely without alternatives but carve out most of the logical space available for the formulation of a precise and consistent quantum theory. Of little interest are the various ``interpretations'' of quantum mechanics that implicitly or explicitly deny at least one of the three assumptions without taking the next step of admitting state variables over and above the wave function, replacing the Schrödinger equation with an equally precise law, or trying to develop a coherent description of nature in which mutually contradictory measurement outcomes coexist. This includes the old Copenhagen-style quantum mechanics that can be understood as choosing options one and two \emph{some of the time} -- insisting that the wave function provides a complete state description for some kinds of systems but not for others, or that the Schrödinger equation is valid for some kinds of interactions but not for others -- with a shifty split somewhere between the ``microscopic'' and the ``macroscopic'' world.

Indeed, if the measurement problem is acknowledged at all from an orthodox perspective, then usually as a challenge to improve upon the ``unprofessionally vague'' (Bell \cite[p. 173]{bell2004}) collapse postulate which posits that whenever a measurement occurs, the Schrödinger evolution is suspended in favor of a probabilistic state reduction. In the spirit of ``interpreting'' quantum mechanics, this call for precision is generally not met by rigorous mathematics -- as in the GRW theory -- but in prose. The idea is that if we could only provide a better answer to Bell's questions, ``what exactly qualifies some physical systems to play the role of `measurer''' \citep[p. 216]{bell2004} and what distinguishes a measurement from other physical interactions, the problem would be solved. 


Aside from hardly meeting the call for professional rigor, such attempts to interpret the measurement problem away are also missing the deeper issue. This issue (which doesn't quite come out with Maudlin's trilemma) is not what to make of a superposed wave function of ``dead cat'' and ``alive cat,'' but what a wave function has to do with a cat in the first place. How, in other words, is the quantum formalism supposed to connect to the world that we experience?

To even begin to answer this question, orthodox quantum mechanics introduces a number of axioms that are at least as problematic as the infamous collapse postulate. If we look at an honest textbook like Cohen-Tannoudji \textit{et al.} \cite{cohen.etal1991}, we find postulates such as their postulates 2-4:
\begin{quotation}
\noindent 
\begin{enumerate}
\setcounter{enumi}{1}
	\item  Every measurable physical quantity $Q$ is described by an operator $\hat{Q}$; this operator is called an observable.
	
	\item  The only possible result of the measurement of a physical quantity $Q$ is one of the eigenvalues of the corresponding observable $\hat{Q}$.
	
	\item  When the physical quantity $Q$ is measured on a system in the normalized state $\psi$, the probability $\mathbb{P}(q_n)$ of obtaining the non-degenerate eigenvalue $q_n$ of the corresponding observable $\hat{Q}$ is
	\begin{equation*}\mathbb{P}(q_n) = \int \phi_n^* \psi \end{equation*}
	where $\phi_n$ is the normalized eigenvector of $\hat{Q}$ associated with the eigenvalue
	$q_n$.
\end{enumerate}
\end{quotation}

\noindent We are not going to dwell on the question of exactly which observable-operator is supposed to correspond to the vital status of a cat. The point is that textbook quantum mechanics relies on problematic postulates about ``measurements'' long before we arrive at the contradiction between the wave function collapse (Cohen-Tannoudji's postulate 5) and the linear Schrödinger evolution (postulate 6); axioms that already require us to accept ``measurements'' as primitive and hardly apply outside of more or less idealized laboratory settings. Even within a laboratory context, it is only testimony to the ingenuity of physicists that the quantum recipe works as well as it does, that in most cases, physicists will figure out what operators to use and what matching experiments to perform, even though it does not -- and cannot -- follow from an \emph{analysis} of the theory. Since the link between ``observable quantities'' and Hilbert space operators is axiomatic, standard quantum mechanics does not have the resources to explain why a certain measurement procedure is associated with a particular observable-operator -- or any operator at all. 

In contrast, Bohmian mechanics allow us to \emph{derive} the operator formalism from a statistical analysis of the measurement process (see \cite[Ch. 3]{duerr.etal2013}, \cite[Ch. 7]{durr.lazarovici2020}). This analysis also clarifies why the observable values are, in general, \emph{produced} rather than \emph{revealed} by measurements, i.e., why they cannot be thought of as physical properties that a system possesses prior to, or independent of, the measurement process. One can see why this leads to all sorts of paradoxes if the observables are conceived as the only, or most basic, link between theory and observable reality (cf. \cite{lazarovici.etal2018}). It is utterly unmysterious if observables can be analyzed in terms of \emph{beables} as Bell insisted:

\begin{quote}
The concept of `observable'  lends itself to very precise \emph{mathematics} when identified with `self-adjoint operator'. But physically, it is a rather wooly concept. It is not easy to identify precisely which physical processes are to be given status of `observations’ and which are to be relegated to the limbo between one observation and another. So it could be hoped that some increase in precision might be possible by concentration on the \emph{be}ables, which can be described in `classical terms' because they are there. The beables must include the settings of switches and knobs on experimental equipment, the currents in coils, and the readings of instruments. `Observables' must be \emph{made}, somehow, out of beables. The theory of local beables should contain, and give precise physical meaning to, the algebra of local observables \citep[p. 52]{bell2004}.
\end{quote}

While standard quantum mechanics reproduces the empirical predictions of Bohmian mechanics whenever its rules are unambiguous, this is not always the case. For instance, \emph{Wigner's friend}-type thought experiments in which observers become a part of measured quantum systems have caused quite the commotion in recent years since plausible applications of the quantum recipe yield inconsistent predictions for the different observers \cite{frauchiger.renner2018}. Such setups are unproblematic for theories like Bohmian mechanics that treat ``observers'' just like any other physical subsystem \cite{lazarovici.hubert2019}.

Closer to feasible experiments is the question of \emph{arrival time distributions}: How long will it take an electron to hit a detector screen after leaving a particle source? It is unclear what operator (or \emph{POVM}, a \emph{positive-operator-valued measure} generalizing the notion of self-adjoint observables) is supposed to be associated with arrival time measurements. There are various contradictory proposals in the literature, none of which seems entirely convincing \cite{vona.etal2013, das.struyve2021}. It helps that Bohm's theory describes localized particles that actually move from the source to the detector. Not only that it makes sense of what is being measured in the first place; it also allows us to compute the distribution of flight times after which the particle trajectories first cross the detector surface instead of guessing for some abstract operator \cite{das.durr2019}.

With the growing understanding that measurements and observables cannot be primitives in the formulation of the theory, modern quantum theories have, by and large, followed two different strategies to provide a precise and objective description of nature (including but not limited to measurement processes). The first is the primitive ontology program with Bohmian mechanics as the prime example. Crucially, PO theories relieve the wave function from the burden of representing matter, its role being instead a dynamical one for the evolution of the PO. The Born rule is then derived from a statistical analysis of the theory, justifying the role of wave functions in making statistical predictions.

The other strategy can be subsumed under the moniker of \emph{quantum state realism}. It tries to develop an objective description of nature by locating three-dimensional objects as patterns in the wave function (see, in particular, Albert's approach \cite{albert2013}) or in a more abstract notion of quantum state (see, in particular, the spacetime state realism of Wallace and Timpson \cite{wallace.timpson2010, wallace2012}). In contrast to the primitive ontology program, the basic relation between the fundamental ontology and manifest macro-objects is then not one of mereological composition but of functional enactment. The modern versions of Everettian quantum mechanics, i.e., Many-Worlds theories, usually fall into this camp. 

We will explain below why this second program remains uncompelling. Here, we want to highlight the key insight behind both approaches that if a theory has well-defined dynamical laws and a clear ontology in terms of which it can describe cats and dogs and pointers on measurement devices, there can be no measurement problem. The description of nature provided by such a theory can be wrong, in that it does not match the empirical facts, but it cannot be paradoxical. In particular, whether measurements result in definite outcomes (as in Bohmian mechanics), or in ``many worlds'' (as in Everettian quantum mechanics), or in some unrecognizable mess, is not \emph{postulated} or \emph{interpreted} but \emph{inferred} by an analysis of the theory. 

Objective collapse theories are generally regarded as a third option for solving the measurement problem but fall into either one of the two camps when it comes to the role of the wave function. The original GRW theory (now sometimes called GRW0) is a theory about the wave function alone. It modifies the linear Schrödinger evolution to suppress macroscopic superpositions like that of ``dead cat'' and ``alive cat.'' But there is still a difference between a cat and the wave function of a cat -- even a collapsed one -- and GRW0 faces the same challenges as Everettian quantum mechanics in making the connection. Nowadays, it is thus common to equip the GRW theory with a primitive ontology. One proposal (due to John Bell \cite[Ch. 22]{bell2004}) regards the collapse centers themselves as the primitive ontology, discrete ``matter flashes'' in space and time that constitute macro-objects (GRWf). Another proposal (due to Ghirardi \textit{et al}. \cite{ghirardi.etal1995}) is called GRWm and uses the  wave function to define a continuous mass density field in physical space 
\begin{equation}\label{mdef}
  m(\vec x,t) = \sum_{i=1}^N m_i \int \mathrm{d}^3 q_1 \cdots  \mathrm{d}^3q_N \,
  \delta(\vec x - \vec q_i)\bigl\lvert\Psi(\vec q_1, \ldots,  \vec q_N,t) \bigr\rvert^2 \,.
\end{equation}
Notably, this PO is also compatible with Many-Worlds, where it makes for a clearer theory than Everettian approaches \citep{allori.etal2011}. In any case, the wave function then assumes a dynamical role for the evolution of the primitive ontology -- much like it does in Bohmian mechanics -- although in the case of collapse theories, this evolution is intrinsically stochastic. 

In the upshot, if the measurement problem is understood as the problem of connecting the theoretical formalism to observable facts, the collapse dynamics add little to the solutions provided by Bohmian mechanics or modern versions of Many-Worlds (cf. Esfeld \cite{esfeld2018}). But since they lead to (in principle) testable effects that distinguish spontaneous collapse from unitary quantum theories, their viability has become a primarily empirical question.


 
 \section{Looking for the World in the Wave Function}\label{sec:WFrealism}
 
  Let us take a  look at approaches that try to provide a description of nature in terms of the wave function or quantum state without involving a primitive ontology. These approaches agree on the idea that one can identify familiar physical objects in the wave function if one looks at it the right way, but they disagree, at times fundamentally, on what the right way of looking is. Here is a list of popular proposals without any claim to completeness: 
  
  \begin{enumerate} 
     \item The wave function realism of David Albert \cite{albert2013, albert2015} (proposed in the context of GRW0 but, in principle, equally compatible with a Many-Worlds interpretation) according to which macro-objects are functionally enacted by three-dimensional projections of the wave function. Albert interprets the wave function as a physical field on a $3N$-dimensional space and calls its projections on three-dimensional space ''shadows''.
     \item Alyssa Ney \cite{ney2017, ney2021} also regards the wave function on $3N$-dimensional space as fundamental but interprets it as grounding ``partial instantiations'' of particle configuration in a (non-fundamental) three-dimensional space.
     \item Lev Vaidman's \cite{vaidman2019} pragmatic interpretation of Everettian quantum mechanics identifies a ``world'' with a special structure of wave function branches in which some of its components are well-localized on the centers of masses of macro-objects in three-dimensional space. 
     \item The \emph{Spacetime State Realism} of Wallace and Timpson \cite{wallace.timpson2010, wallace2012} associates a quantum state to spacetime regions, taking four-dimensional spacetime as fundamental. The physical properties instantiated in those regions are supposed to be described by an algebra of local operators as used in algebraic formulations of QFT.\footnote{Seemingly based on the questionable idea that they are better candidates for ontology than the observable-operators of non-relativistic quantum mechanics.}
    \item The \textit{Mad-Dog Everettianism} of Carroll and Singh \cite{carroll.singh2019} regards Hilbert space as fundamental and tries to extract empirical content from the spectrum of the Hamiltonian. 
    \item We can also mention Everett's \emph{relative state formalism} \cite{everett1957}, although it is rarely taken seriously anymore as it still relies on observable-operators and a decomposition of the wave function into corresponding eigenstates.
\end{enumerate}
The mere existence of these many contradictory interpretations should disabuse us of the often propagated idea that there is anything like a ``straightforward'' or ``literal'' reading of the wave function formalism. The ongoing controversies about whether and how quantum theories without PO make contact with observable reality are rather evidence that these theories are simply incomplete. 

For the most instructive comparison with the primitive ontology approach, we will focus on the program of wave function realism as exemplified, in particular, by Albert's functionalism. One can understand the idea that the wave function, in either a spontaneous collapse or Many-Worlds framework, should be able to ``enact'' (what appears as) localized objects in 3-dimensional space. Also in Bohmian mechanics, the particle dynamics are encoded in the wave function. And when a collection of particles forms a macro-object that behaves like a semi-classical body, the respective subsystem will be guided by an effective wave function that is well-localized in configuration space and whose peak propagates on an approximately Newtonian trajectory.\footnote{For the definition of effective wave functions of subsystems, see \cite{duerr.etal1992}. For the classical limit of Bohmian mechanics, see \cite{allori.etal2002}, reprinted as Ch. 5 in \cite{duerr.etal2013}, or the textbook of Dürr and Teufel \cite{durr.teufel2009}.} This can lead one to think that the particles are dispensable, that all there is to there being a table or a cat or a measurement device are wave packets evolving a certain way. 

The idea has some prima facie plausibility if one is used to thinking of the wave function on \emph{configuration space}. But a \emph{configuration space} only makes sense given an ontology on three-dimensional space whose spatial degrees of freedom are mathematically represented on something like $\mathbb{R}^{3N}$. As de Broglie already said in his report to the 1927 Solvay Conference ``it seems a little paradoxical to construct a configuration space with the coordinates of points that do not exist'' \cite[p. 346]{bacciagaluppi.valentini2007}.

Taking wave function realism seriously is to take $3N$-dimensional space seriously as the fundamental stage of physics. A priori, there is then no reason why coordinate-triplets in this high-dimensional space should refer to points in some three-dimensional space, and no fact as to \emph{which} coordinate-triplets jointly designate such a point \cite{monton2002, monton2006}. Even more devastating to the functionalist program, different coordinate-triplets of points in $3N$-space do not bear any metric relation to one another. There is no meaningful distance between internal degrees of freedom of the wave function -- e.g., between wave packets living in different subspaces of $\mathbb{R}^{3N}$ -- any more than between the $x,y$, and $z$ coordinates of your hand. Thus, even if we grant that all there is to being ``a table or a chair or a baseball or an observer ... is to occupy this or that particular \emph{niche} in the causal map of the world'' \citep[p. 132]{albert2015}, patterns in the wave function are incapable of occupying those niches. At least if the relevant causal relations have anything to do with how things \emph{move} depending on their relative positions. 

Bohmians are also ``macro-object functionalists'' in the obvious sense that a table-shaped configuration of particles must have the right dynamics in order to realize a \emph{table}. But while it is fairly obvious how macro-objects moving and interacting in physical space could be functionalized in terms of other things moving and interacting in physical space, it is hopeless to provide a functional definition of all of three-dimensional reality in the language of the wave function, eliminating (as one would have to) any reference to three-dimensional geometry and matter located in it. 

The best continuation of the wave function realist program is to ``recover'' (or, to put it more cynically, smuggle in) the structure of a $N\times 3$-dimensional configuration space by arguing that it makes the wave function dynamics particularly simple, or informative, or symmetric (see, in particular, Ney \cite{ney2015}). But it is not clear to us what such arguments could accomplish. If one wants to draw ontological inferences from the $N\times 3$-dimensional structure ``hiding'' in the wave function, the best inference by far is to a primitive ontology of $N$ particles in three-dimensional space. If wave function realists are merely making the case for a convenient mathematical representation in which to look for patterns in $\Psi$, they would seem to suggest that observable reality is somehow an efficient summary of their pet theory -- which gets the point of the whole exercise (i.e., science) pretty much backwards. 

In any case, when all is said and done, Vaidman refers to three-dimensional macro-objects that have no basis in his theory, Ney talks about particle configurations that could and should have been postulated from the beginning, and Albert's  ``shadows'' that are supposed to ``formally enact'' three-dimensional objects are simply fields 
\begin{equation}\label{shadows} f_i(\vec x)  := \int \mathrm{d}^3 q_1 \cdots  \mathrm{d}^3q_N \,
  \delta(\vec x - \vec q_i)\bigl\lvert\Psi(\vec q_1, \ldots,  \vec q_N) \bigr\rvert^2, \; i= 1,...,N 
\end{equation}
on three-space, obtained as one-particle marginals of the $\lvert\Psi\rvert^2$-density. The main difference to \eqref{mdef} is that we now have $N$ fields that can overlap rather than a single mass-density defined as their weighted sum.\footnote{That the projections are not multiplied by constants $m_i$ is secondary; the masses appear as parameters in the Schrödinger equation whence they affect the dynamics of the shadows.} For all practical purposes, these shadow fields \emph{are a primitive ontology} -- not the only point that would be much clearer if \eqref{shadows} was introduced as a defining equation of the theory rather than a strategy for reading something into the wave function.

We can only briefly discuss Albert's reason for regarding the ``shadows'' as merely derivative mathematical projections and insisting that the fundamental stage of quantum theory is the high-dimensional space on which the wave function is defined (a stance that he adopts even with respect to Bohmian mechanics, see his ``marvelous point'' interpretation in \cite{albert2015}). Albert is concerned about the so-called ``problem of connection,'' simply put, the question how the wave function could have anything to do with the dynamics of a PO if they live in different spaces -- the $3N$-dimensional ``configuration'' space versus the three-dimensional ``physical'' space. Various answers have been proposed in the philosophical literature, from nomological interpretations of the wave function (e.g., \cite{esfeld.etal2014}), to regarding it as a ``multi-field'' in three-dimensional space \cite{hubert.romano2018}, to simply accepting the quantum state as a non-local feature of (three-dimensional) reality that doesn't fall into any of the classical ontological categories \cite{maudlin2019}. One can find reasons to dislike all of these answers. The wave function is, in any case, a strange kind of object.\footnote{Which only underscores the need to understand its physical role in terms of a PO.} But tables and chairs and baseballs are familiar objects, and they are certainly \emph{not} mathematical projections. Thus, if one cares about distinguishing abstract mathematical constructions from what is ``real'' according to the theory, the latter better include the things that stones and tables and cats are made out of or otherwise realized by. There is no end to possible mathematical projections (and other mappings) that can be defined within a theory, but they don't yield anything \emph{physical}, as Maudlin's example forcefully demonstrates: 

\begin{quote}
Consider a regular low-dimensional Newtonian world with
tables and chairs and baseballs all composed of particles. And
now define the ``3-foot north projection'' of any particle to be the
point in space exactly three feet to the north (i.e. in the direction
from the center of the earth to the center of Polaris) of the
location of the particle. Then trivially the 3-foot north projections
of all the particles in a table will be a set of locations that have
the same geometrical structure as the particles in the table. And
the 3-foot north projections of all the actual particles in tables
and chair and baseballs will formally enact, in Albert’s sense, the
tables and chairs and baseballs and observers whose projections
they are. But these ``formal enactments'' are clearly not tables and
chairs, and the 3-foot north projection of a person having a headache
is clearly not an actual sentient person with a headache. \cite[p. 126]{maudlin2019}
      \end{quote}
      
There are no purely formal features (such as being defined on a three-dimensional space) that alone qualify a mathematical structure to represent something \emph{physical}. Some ontological commitment(s) must be part of the physical theory. This can include an ontological commitment to the wave function or quantum state, but it won't be enough to describe a world that looks even remotely like ours. Given that all our empirical evidence is ultimately evidence of localized matter in space and time, the most natural way to account for it is in terms of a primitive ontology. And all claims that quantum mechanics compels us to look for less natural ways -- or give up on an objective description of nature altogether -- were proven wrong by Bohm. 
    
\section{Einstein's Epistemological Model}
The point of a primitive ontology goes deeper than accounting for empirical evidence in the simplest and most compelling fashion. It addresses more fundamental questions about the relation between theory and experience. How can a theoretical formalism make contact with our experience in the first place? How do theoretical concepts, removed from direct sense experience, acquire meaning? Such questions have not been the exclusive concern of professional philosophers. One of the most insightful, though brief, reflections on the matter can be found in Einstein's ``autobiographical notes'' in \cite{schilpp1949}.

\begin{quote}
    I see on the one side the totality of sense experiences, and, on the other, the totality of the concepts and propositions which are laid down in books. The relation between the concepts and propositions among themselves and each other are of a logical nature, and the business of logical thinking is strictly limited to the achievement of the connection between concepts and propositions among each other according to firmly laid down rules, which are the concern of logic. The concepts and propositions get ``meaning,'' viz., ``content,'' only through their connection with sense experiences. The connection of the latter with the former is purely intuitive, not itself of a logical nature. The degree of certainty with which this relation, viz., intuitive connection, can be undertaken, and nothing else, differentiates empty phantasy from scientific ``truth''. 
    \citep[pp. 11-13]{schilpp1949}
\end{quote}

\begin{figure}[ht]
    \centering
   \includegraphics[width=\textwidth]{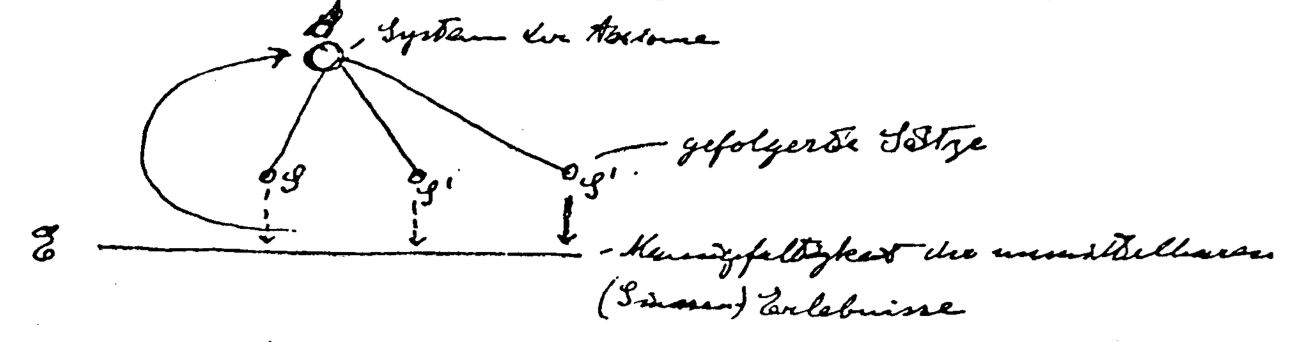}
    \caption{Einstein's epistemological model sketched by Albert Einstein in his letter to Maurice Solovine from May 7, 1952. Graphic adopted from \cite{einstein2011}. }
    \label{fig:Einsteinmodel}
\end{figure}

\noindent Fortunately, Einstein further explained his view in a letter to his friend Maurice Solovine (from May 7, 1952, reprinted in \cite{einstein2011}) in which he (literally) sketches his epistemological model (Fig. \ref{fig:Einsteinmodel}). 

\noindent We quote the relevant passage in full: 

\begin{quote}
\begin{enumerate}[1)]\small
 \item The $\mathcal{E}$’s (experiences) are given to us.
\item The axioms from which we draw our conclusions are indicated by $\mathcal{A}$. Psychologically the $\mathcal{A}$’s depend on the $\mathcal{E}$’s. But there is no logical route leading from the $\mathcal{E}$’s to the $\mathcal{A}$’s, but only an intuitive (psychological) connection, which is always subject to revision.
\item From $\mathcal{A}$, specific statements $\mathcal{S}$, $\mathcal{S}'$, $\mathcal{S}''$ are logically deduced; these deductions can lay claim to exactness.
\item The $\mathcal{S}$’s are connected to the $\mathcal{E}$’s (verification through experience). Closer examination shows that this procedure also belongs to the extralogical (intuitive) sphere, for the relation between the terms appearing in $\mathcal{S}$ and the immediate experiences is not logical in nature.\\
But the relation between $\mathcal{S}$’s and $\mathcal{E}$’s is (pragmatically) much less uncertain than the relation between the $\mathcal{A}$’s and the $\mathcal{E}$’s. (Take the notion ``dog'' and the corresponding immediate experiences.) If such a relationship could not be set up with a high degree of certainty (though it may be beyond the reach of logic), logical machinery would have no value in the ``comprehension of reality'' (example: theology).
\end{enumerate}

\noindent This all comes down to the eternally problematical connection between that which is thought and that which can be experienced (sense experience). \citep[pp. 123-124; translation corrected by the authors.]{einstein2011}
\end{quote}

\noindent The point Einstein is making in (4) is the following: Nothing about our sense experience follows deductively from a physical theory (barring a solution to the hard problem of consciousness). There is always a logical gap between the singular statements, i.e., predictions, that can be derived from the theory and the immediate sense experience that these predictions are tested against. Up to this point, Einstein's diagnosis does not differ much from that of logical empiricists like Hempel \cite{hempel1966}, Carnap \cite{carnap1958}, or Nagel \cite{nagel1961}, who tried to employ additional postulates -- \emph{bridge principles} -- to translate theoretical statements into observational ones. But Einstein's schema has no room for further postulates in between the theoretical statements $\mathcal{S}$ and the sense experiences $\mathcal{E}$. He insists that this connection, though ``extralogical,'' must be more or less self-evident so it can be intuited with ``a high degree of certainty.'' 

This does not mean that a theory cannot make counter-intuitive predictions (Einstein's theories certainly do), nor that it can't be hard to derive them. What has to be intuitive is the relation between derived theoretical statements and their empirical content. Otherwise, the ``logical machinery would have no value in the `comprehension of reality'''.

The concept of primitive ontology is very much in line with Einstein's epistemological model. Statements about the primitive ontology derived from the theory -- in situations where configurations of the primitive entities form macroscopic bodies -- provide the points of closest contact between theory and experience. The remaining gap is small enough to be bridged by somewhat sophisticated intuition (e.g., sophisticated enough to understand \emph{atomism}). There is an obvious enough connection between the physical objects populating the world we experience and suitable configurations of point particles or GRW \emph{flashes} or a mass-density field. There is no need for additional rules to translate between ``theoretical language'' and ``observational language'' because both speak of configurations of matter in space and time.

It is instructive to compare a Bohmian prediction for, say, pointer positions with other propositions that can be derived from its mathematical formalism but do not refer to the PO. Take the proposition: \emph{The Coulomb Hamiltonian is essentially self-adjoint}. As a mathematical statement, it is perfectly clear. But by itself, this statement has no empirical content and no value in the comprehension of physical reality. It gets physical meaning only through the way it pertains to particle dynamics and the statistics of particle positions -- and thereby connects to concrete facts in the world. 

Returning to Einstein's scheme, we understand that the connection between the $\mathcal{S}$'s and the $\mathcal{E}$'s, between certain singular statements of the theory and observable facts in reality, must be ``intuitive'' precisely because it is the one (and final) step on the path from the theory to experience that cannot be subject to physical or mathematical analysis.  
One of the most fundamental problems of quantum mechanics -- the most basic sense in which the theory is incomplete -- is that it leaves a huge conceptual and epistemological gap between its mathematical formalism and observable reality. (In line with the Copenhagen dogmatism that declared the quantum realm to be ``unanschaulich'' and regarded the formalism as a ``purely symbolic scheme'' \citep[p. 210]{schilpp1949} devoid of physical meaning.)

Textbook quantum mechanics with its accompanying measurement postulates is a prime example of what goes wrong if one tries to bridge this gap by additional axioms that are as tentative and abstract as the intra-theoretical ones -- only much more vague. The path from the $\mathcal{A}$'s to $\mathcal{E}$'s, including the measurement process, is almost entirely removed from physical analysis, so that it remains ambiguous, non-explanatory, and by design incapable of providing a coherent image of reality. 

The various versions of wave function or quantum state realism are examples (though by far not the worst) of trying to bridge the gap to observable reality by roundabout ``interpretations,'' as if it could be left up to philosophical speculations to read empirical content into a physical theory. At best, as we have argued, these approaches are primitive ontology theories in denial. At worst, they are unintelligible. In any case, they are working from the end backward, relying on arbitrary constructions and hand-waving that hardly inspire a high degree of certainty about what a world described by the theory would actually look like. 

Well-intentioned but equally misguided is the idea that connecting quantum theory to the world is merely a matter of providing the right mathematical formulation or investigation. The question of how a mathematical formalism connects to the world cannot be answered by more or different mathematics. It has to be clear from the onset, from the first principles of the theory, as is the case when it starts with a primitive ontology. Then and only then is extracting its empirical content merely a matter of analyzing the theory. 

Through its primitive ontology of particles, Bohmian mechanics provides a clear and coherent image of the physical world, one that is not too difficult to integrate with our manifest image, at least not more than necessary. It thereby clarifies not only the meaning of other theoretical structures figuring into its fundamental laws but also of the standard quantum formalism that can be derived from them. 

Regarding the general value of such a coherent spatio-temporal image of nature, one may go as far as to claim that it not only allows for a better, less ambiguous and more explanatory way of doing physics but is really the point of the entire enterprise. As Schrödinger wrote in a letter to Wilhelm Wien (on August 25, 1926):
``Bohr's point of view that a spatio-temporal description is impossible, I reject \emph{a limine}. Physics is not all atomic research, science is not all physics, and life is not all science. The purpose of atomic research is to integrate the pertinent experiences with the rest of our thinking. All this other thinking, as far as it concerns the outside world, is situated in space and time. If the embedding into space and time does not succeed, then the whole point is missed, and one does not know which purpose should be served with it at all.'' \cite[p. 306; translation by the authors]{vonmeyenn2011}

\newpage 
\backmatter
\bmhead{Acknowledgments}
We thank Christian Beck for valuable discussions. Nino Zanghì was the first to tell us about Einstein's epistemological model. About the epistemological role of primitive ontology, we learned a lot from Tim Maudlin's talk ``How Theory Meets the World'' that we had the pleasure of attending many years ago. Section 3 of this essay was adapted from the forthcoming book ``Typicality Reasoning in Probability, Physics, and Metaphysics'' by Dustin Lazarovici.

\newpage

\bibliography{sn-bibliography}

\end{document}